\documentclass[11pt]{article}
\usepackage{amssymb}
\usepackage{graphics,epsfig,subfigure}
\usepackage{diagbox}
\usepackage{amsthm}
\usepackage{amscd}
\usepackage{amstext}
\newcommand{\ba}{\begin{array}}
\newcommand{\ea}{\end{array}}

\let\w=\omega

\def\be{\begin{equation}}
\def\ee{\end{equation}}
\def\ba{\begin{array}}
\def\ea{\end{array}}

\def\dalemb#1#2{{\vbox{\hrule height .#2pt
        \hbox{\vrule width.#2pt height#1pt \kern#1pt
                \vrule width.#2pt}
        \hrule height.#2pt}}}

\def\ocal{{\mathcal{O}}}

\begin{document}

\begin{center}


\vspace{1cm} { \LARGE {\bf Excited states of holographic superconductors}}

\vspace{1.1cm}

Yong-Qiang Wang\footnote{yqwang@lzu.edu.cn, corresponding author}, Tong-Tong Hu\footnote{hutt17@lzu.edu.cn}, Yu-Xiao Liu\footnote{liuyx@lzu.edu.cn}, Jie Yang\footnote{yangjiev@lzu.edu.cn}
 and Li Zhao\footnote{lizhao@lzu.edu.cn}

\vspace{0.7cm}

{\it Institute of Theoretical Physics $\&$ Research Center of Gravitation, Lanzhou University, Lanzhou 730000, People's Republic of China }

\vspace{1.5cm}

\end{center}

\begin{abstract}
\noindent
In this paper we re-investigate the model of the
anti-de Sitter gravity coupled to  Maxwell
 and charged scalar fields,  which has been  studied as the gravitational dual to a superconductor  for a long time since the famous work [Phys.\ Rev.\ Lett.\  {\bf 101}, 031601 (2008)]. By numerical method, we present a novel family of solutions of holographical superconductor with excited states,
and find  there exists a lower
critical temperature in the corresponding excited state.
Moreover, we study the condensate and  conductivity in the excited states.
 It is very interesting that the conductivity $\sigma$ of each excited state has an additional pole in
$\text{Im}[\sigma]$ and a delta function in $\text{Re}[\sigma]$  arising at the low temperature inside the gap,
which   is just the evidence of the existence of  excited states.
\end{abstract}

\vspace{5cm}

\pagebreak

\section{Introduction}
In condensed matter physics, there is still no consensus on the mechanism of  high temperature superconductivity.
 Over the past decade,
 the AdS/CFT correspondence \cite{Maldacena:1997re,Witten:1998qj,Aharony:1999ti} is one of the most important results from string theory, which establishes the relationship between the strongly correlated
field on the  boundary and a weak gravity theory
in one higher-dimensional   bulk spacetime.
In the past ten years, the AdS/CFT correspondence has provided  a novel way to study
condensed matter theory, and received a great deal of attention.
In the seminal papers \cite{Gubser:2008px,Hartnoll:2008vx,Hartnoll:2008kx}, the authors investigated the model of a
complex scalar field coupled to a $U(1)$ gauge field in a (\(3+1\))-dimensional Schwarzschild-AdS
black hole and found that due to the
$U(1)$ symmetry breaking below the critical temperature $T_c$, the condensate of the scalar field could be
interpreted as the Cooper pair-like superconductor condensate. Moreover, there exists a  gap in the optical conductivity of the superconducting state,
and the value of the  gap
 is close to the value of a high temperature superconductor. Thus,
this model can  be regarded as the dual explanation to the high temperature
superconductor.
When replacing the scalar field with other matter fields,  one can also obtain the condensate of   matter fields corresponding to various kinds of holographic superconductors.
For example,
holographic d-wave model was constructed
by introducing  a symmetric, traceless second-rank tensor coupled to a $U(1)$ gauge field  in the bulk \cite{Chen:2010mk,Benini:2010pr,Kim:2013oba}.
With the SU(2) Yang-Mills field coupled
to gravity, the holographic p-wave model has also been discussed
in \cite{Gubser:2008wv}.
Two alternative holographic realizations of p-wave superconductivity  could arise from the condensate of a two-form field \cite{Aprile:2010ge} and  a complex,
massive vector field with $U(1)$ charge \cite{Cai:2013pda,Cai:2013aca}, respectively.
The model of  holographic superconductor can  also be extended to study the holographic Josephson junction \cite{Horowitz:2011dz,Wang:2011rva,Siani:2011uj,Kiritsis:2011zq,Wang:2011ri,Wang:2012yj,Cai:2013sua,Takeuchi:2013kra,Li:2014xia,Liu:2015zca,Hu:2015dnl,Wang:2016jov,Kiczek:2019lmz}, which is
 made up of two superconductor
materials with weak link barrier \cite{Josephson:1962zz}. A top-down construction of holographic superconductor from superstring theory was discussed in \cite{Gubser:2009qm}, and a similar construction using an M-theory truncation was proposed in \cite{Gauntlett:2009dn, Gauntlett:2009bh}.
For reviews of holographic superconductors,
see \cite{Hartnoll:2009sz,Herzog:2009xv,Horowitz:2010gk,Cai:2015cya}.

Until now, lots of work of holographic superconductor were investigated in
the ground state\footnote{In  \cite{Gubser:2009cg,Horowitz:2009ij} the ground state represents the zero-temperature limit of holographic superconductor while in this paper the ground state refers to the scalar field without nodes.}, that is, the scalar
field can keep sign along the radial  direction.
It is well-known that the excited states could have some nodes  along the radial  direction,  where  the value  of the scalar field could change the sign.
On the other hand,  it is easy to see in  quantum theory that the bound states  with
given angular momentum and other quantum numbers  similarly could form towers of states known
as radial excitations with otherwise identical quantum numbers. Especially, holographic phenomenological models  \cite{Erlich:2005qh,Karch:2006pv} have the potential to provide a better understanding of strongly interacting systems of quarks and gluons, including the excited states of  hadrons. These models are presently known as AdS/QCD. Furthermore, in the background of Schwarzschild black hole, holographic  model of QCD could be extended to the finite temperature system \cite{Ghoroku:2005kg},
There  are  also many new progresses on
the excited states of  hadrons from holographic QCD,  such as \cite{Afonin:2009xi,Fujita:2009wc,Cui:2011ag}.

So, it is interesting to see whether there exist the solutions
of holographic superconductor with excited states of the scalar field. In the present paper, we
would like to numerically solve the
Maxwell-Klein-Gordon system with the background of a four-dimensional Schwarzschild-AdS
black hole and give a family of
excited states of holographic superconductors  with  different critical temperatures.
  Moreover, we will also study the condensate and  optical conductivity in the excited states of holographic superconductors.

The paper is organized as follows: in Sect. \ref{sec2}, we review the model of a   $U(1)$ gauge field coupled with a    charged scalar field in (\(3+1\))-dimensional AdS spacetime and show a gravity dual model of  holographic superconductor. We show the numerical results of the excited states and study the characteristics of the condensate and  conductivity in Sect. \ref{sec3}. The conclusion and discussion are given in the last section.

\section{Review of holographic superconductors}\label{sec2}
Let us introduce the model of a Maxwell field and a charged complex scalar field in the four-dimensional Einstein gravity spacetime with a negative cosmological constant. The bulk action reads
\be
\mathcal{S}= \frac{1}{16\pi G}\int \mathrm{d}^4x \left[R+\frac{6}{\ell^{2}}-\frac{1}{4}F^{\mu\nu}F_{\mu\nu}-(\mathcal{D}_\mu \psi)(\mathcal{D}^\mu \psi)^*-m^{2}\psi \psi^*\right],\label{Lagdensity}
\ee
where $F_{\mu\nu}=\partial_{\mu}A_{\nu}-\partial_{\nu}A_{\mu}$ is the field strength of the $U(1)$ gauge field, and $\mathcal{D}_\mu=\nabla_\mu-iq A_\mu\psi$  is
the gauge covariant derivative with respect to $A_{\mu}$.
The constant $\ell$  is the AdS length scale, $m$ and $q$ are the mass and  charge of the complex scalar field $\psi$, respectively.
Due to the existence of  Maxwell and complex scalar field, the strength of the backreaction of the matter fields on the spacetime metric  could be tuned by the charge $q$.
In order to see that how the effect of  backreaction varies with the charge $q$, we could introduce the scaling transformations $A\rightarrow A/q$ and $\psi\rightarrow \psi/q$, and the Lagrangian density in Eq.
(\ref{Lagdensity}) changes into
\begin{eqnarray}\label{Lagdensity1}
16\pi G \mathcal{L}&=&R+\frac{6}{\ell^{2}}+\kappa \left(-\frac{1}{4}F^{\mu\nu}F_{\mu\nu}-|\nabla\psi-iA\psi|^{2}-m^{2} \psi \psi^*\right),
\end{eqnarray}
with the constant $\kappa\equiv1/q^{2}$. From the above, we can see that the value of the parameter  $\kappa$ will decrease when $q$ increases, and the  backreaction of matter fields
on the spacetime metric could be ignored in the large $q\rightarrow \infty$ limit, which is also called as  the probe limit.
Here we adopt to
the probe approximation. Thus,
the following equations of the scalar and Maxwell fields can be derived from  the Lagrangian density (\ref{Lagdensity1}):
\begin{eqnarray}
(\nabla_{\mu}-iA_{\mu})(\nabla^{\mu}-iA^{\mu})\psi-m^{2}\psi&=&0,\label{scalarequ}\\
\nabla_{\mu}F^{\mu\nu}-i[\psi^{\ast}(\nabla^{\nu}-iA^{\nu})\psi-\psi(\nabla^{\nu}+iA^{\nu})\psi^{\ast}]&=&0.\label{maxwellequ}
\end{eqnarray}
In addition, with  neglect of the backreaction of matter fields on the metric, the solution of Einstein equations is the well-known Schwarzschild anti-de Sitter black hole. The solution with a planar symmetric horizon can be written as follows
\be \label{metric}
  ds^{2} = -f(r)dt^{2}+\frac{dr^{2}}{f(r)}+r^{2}(dx^{2}+dy^{2}),
\ee
where $f(r)=\frac{r^{2}}{\ell^{2}}(1-r_{h}^{3}/r^{3})$,
and $r_{h}$ is the radius of the black hole's event horizon. The Hawking temperature is given by
\be\label{T}
T=\frac{1}{4\pi}\frac{df}{dr}\bigg|_{r=r_{h}}=\frac{3r_{h}}{4\pi \ell^{2}},
\ee
 which can be regarded as the temperature of  holographic superconductors.

In order to build a holographic model of  holographic superconductors, one could introduce
the following ansatz of matter fields:
\begin{equation}\label{ansatz}
  A=\phi(r) dt, \;\;\;\;\; \psi=\psi(r).
\end{equation}
With the above ansatz, the equations of motion for the  scalar field $\psi(r)$ and electrical scalar potential $\phi(r)$ in the
background of the Schwarzschild-AdS black hole are
\begin{eqnarray}
\psi'' + \left(\frac{f'}{f} + \frac{2}{r}\right) \psi' +\frac{\phi^2}{f^2}\psi - \frac{m^2}{\ell^2 f} \psi &=& 0 \,,\label{eom1}\\
\phi'' + \frac{2}{r} \phi' - \frac{2\psi^2}{f} \phi &=& 0.\label{eom2}
\end{eqnarray}
At the AdS boundary, the asymptotic behaviors of the functions $\psi(r)$ and  $\phi(r)$ take the following forms
\begin{eqnarray}\label{asympphi}
\psi &=& \frac{\psi^{(1)}}{r^{\Delta_-}} + \frac{\psi^{(2)}}{r^{\Delta_+}} + \cdots \,,\\
\phi &=& \mu - \frac{\rho}{r} + \cdots \,,
\end{eqnarray}
with
\begin{equation}\label{delta1}
\Delta_\pm=\frac{3\pm\sqrt{9+4 m^2}}{2},
\end{equation}
where the constants $\mu$ and $\rho$ are the chemical potential and charge density in the dual field theory, respectively.
 According to AdS/CFT duality, $\psi^{(i)} (i=1,2) $ are the corresponding expectation values of the dual scalar operators $\mathcal{O}_{i}$, respectively.
Note that  in  the four-dimensional spacetime  the values of mass need to satisfy the Breitenlohner-Freedman (BF) bound of $m^{2}\geq-9/4$ \cite{Breitenlohner:1982bm}.
 In this paper, we will set $m^2=-2$ for simplify.
\section{Numerical results}\label{sec3}
In this section, we will solve the above coupled equations (\ref{eom1}) and (\ref{eom2})  numerically. It is convenient to change the radial coordinate $r$ to  the new radial coordinate $z=r_h/r$.  For simplify, we set $\ell=1$ and $r_h=1$. Thus the inner and outer boundaries of
the shell are fixed at $z = 0$ and $z= 1$, respectively.
All numerical calculations are based on the spectral methods. Typical grids used have the
sizes from $50$ to $300$ in the integration region $0 \leq z \leq 1$. Our iterative process is
the Newton-Raphson method, and the relative error for the numerical solutions in this work is
estimated to be below $10^{-5}$.

In an approach based on the Newton-Raphson method, a good initial guess for the profile of the functions $\psi$ and $\phi$ is an essential condition for a successful implementation.  To obtain numerically  the ground state solution,  one can use the profile of a constant as a initial guess for the function $\psi$. Meanwhile, for the $n$-th excited state case, one need choose a initial guess with $n$ nodes for the function $\psi$.

It is well-known that there exists  a critical temperature $T_{c}$, above which there is no scalar
condensate in the Schwarzschild-AdS black hole background, while for $T\leq T_{c}$, the scalar
condensate begins to appear due to  the  spontaneously broken  $U(1)$ gauge symmetry. Until now, only the ground state of holographic superconductor with a fixed critical temperature was found out, that is, the scalar
field $\psi$ could keep sign along the radial direction. By numerically solving the same equations of motion with boundary conditions as the case of ground state, we could also obtain the solutions of  the excited states, in which the value of the scalar field can change
sign along the radial direction.

\begin{figure}
\begin{center}
	\includegraphics[height=.24\textheight,width=.33\textheight, angle =0]{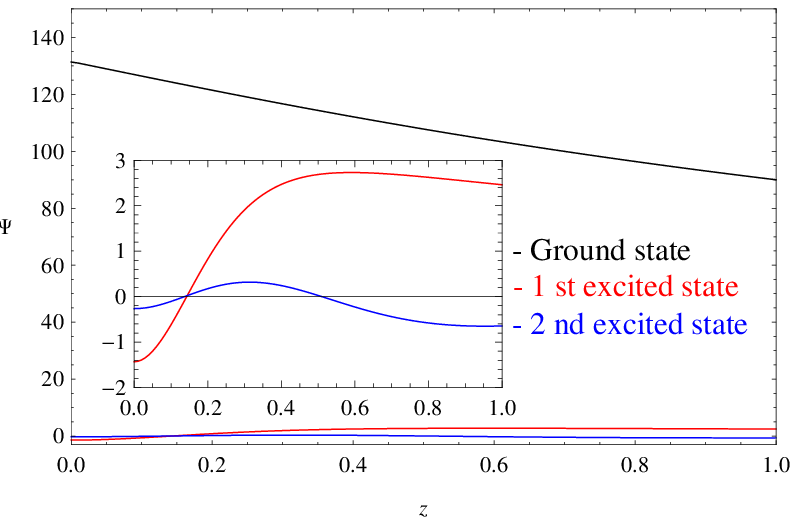}
	\includegraphics[height=.245\textheight,width=.33\textheight, angle =0]{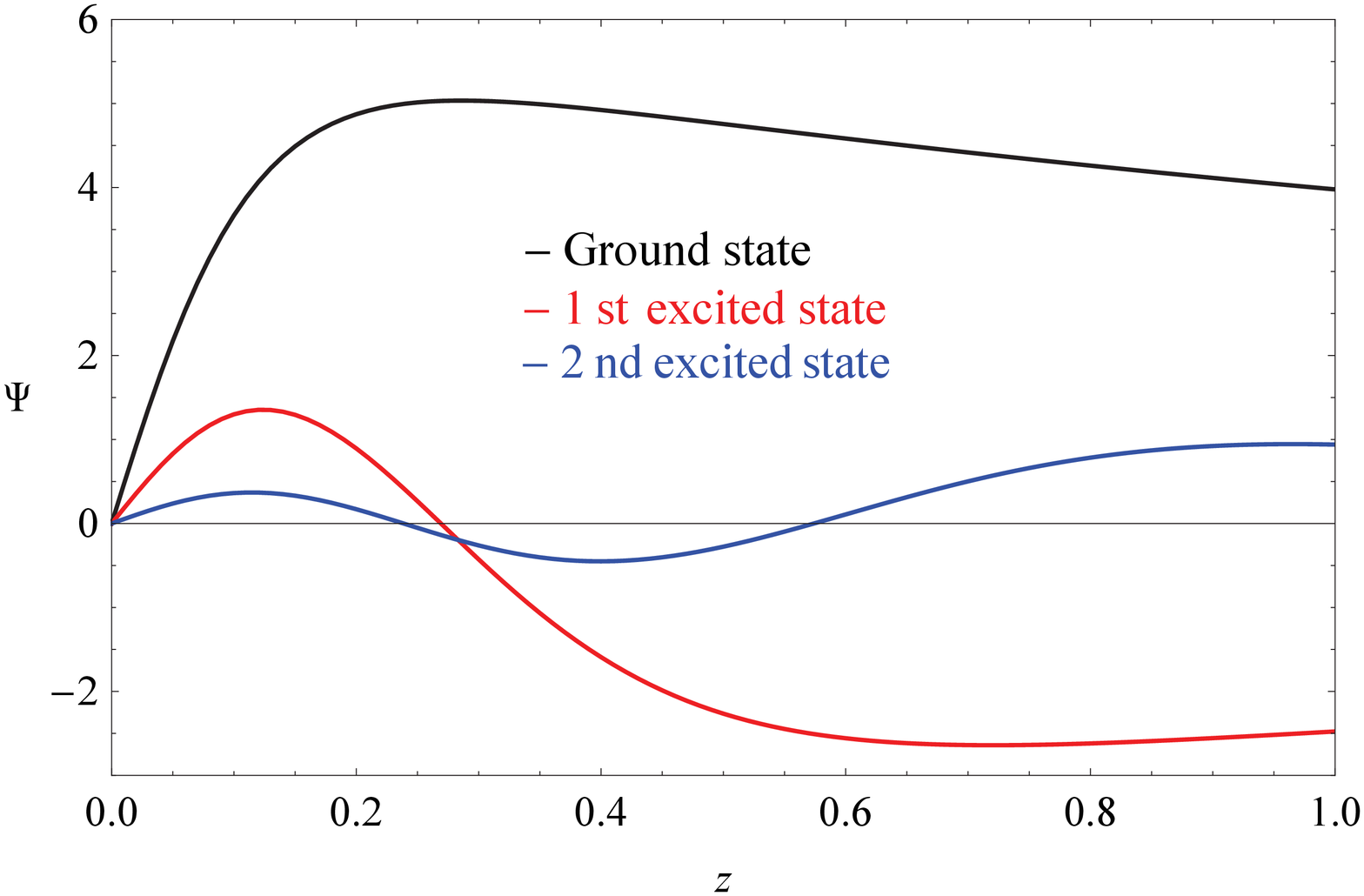}
\end{center}
	\caption{ \textit{Left}: The distribution of the scalar field $\psi$ as a function of  $z$ coordinate  with the chemical potential $\mu=12$   for the condensate $\langle\ocal_1\rangle$.                                                                                                                                                                                               \textit{Right}: The distribution of the scalar field $\psi$ as a function of  $z$ coordinate  with  $\mu=15$   for the condensate $\langle\ocal_2\rangle$.  In both plots, the black,  red  and  blue lines denote ground state, first and second states , respectively.}
	\label{kstability}
\end{figure}
In Fig. \ref{kstability}, we show two kinds of typical
results of the distribution of  the scalar field  $\psi$ as a
function of $z$ for the condensates $\langle\ocal_1\rangle$ and $\langle\ocal_2\rangle$, respectively. The values of the chemical potential $\mu$ are equal to 12 (left panel) and 15 (right panel).
In both plots, the black,  red  and  blue lines denote ground state, first and second states, respectively. The inset in left panel shows the detail of the  curves of the first and second excited states. From Fig. \ref{kstability}, we can see that the ground state has no nodes,  and $n$-th excited state has exactly $n$ nodes, where  the values  of the scalar field are equal to zero. Moreover, the amplitude of $\psi$ deceases with higher excited states.

	\begin{table}[!htbp]
		\centering
		\begin{tabular}{|c|c|c|c|c|c|c|c|c|}
			\hline
       \diagbox{$\;$}{$\mu_{c}$}{n} &$0$&$1$&$2$&$3$&$4$&$5$&$6$\\
            \hline
			$\langle\ocal_1\rangle$ &1.120&6.494&11.701&16.898& 22.094&27.290& 32.486 \\
			\hline
			$\langle\ocal_2\rangle$ &4.064&9.188&14.357&19.538& 24.725& 29.915& 35.107\\
			\hline
		\end{tabular}
\caption{Critical chemical potential $\mu_{c}$ for the operators $\ocal_1$ and $\ocal_2$ from the ground state to sixth excited state.}\label{table1}
	\end{table}

In the table \ref{table1}, we present the
results of the critical chemical potential $\mu_{c}$ for the operators $\ocal_1$ and $\ocal_2$ from the ground state to sixth excited state. We can see that an excited state   can be regarded as a new static solution of holographic  superconductors that has a higher critical chemical potential than the ground state, that is,  an excited state has a higher critical charge density $\rho_c$. Because  the critical temperature is proportional to $\rho^{-1/2}_c$ in four-dimensional spacetime, the excited state has a lower  critical temperature than the ground state.
With the decrease of the temperature,  the ground state first appears, and when the temperature is lower to the critical temperature of the first-excited state,
the solutions of the first-excited state begin to appear.  It is interesting to note that
the difference of critical chemical potential $\mu_{c}$ between the consecutive states is  about $5$ regardless of the condensates $\langle\mathcal{O}_{1}\rangle$ or $\langle\mathcal{O}_{2}\rangle$.
The relation between  $\mu_c$ and $n$   can be fitted as
\begin{displaymath}
\mu_c \approx\left\{ \begin{array}{ll}
5.217 n+ 1.217 \,,  & \qquad\textrm{for} \;\;  \ocal_1,\\
5.177 n+ 4.026 \,,  & \qquad\textrm{for} \;\; \ocal_2.
 \end{array} \right.
\end{displaymath}

\subsection{Condensates}
\begin{figure}[h]
\begin{center}
\epsfig{file=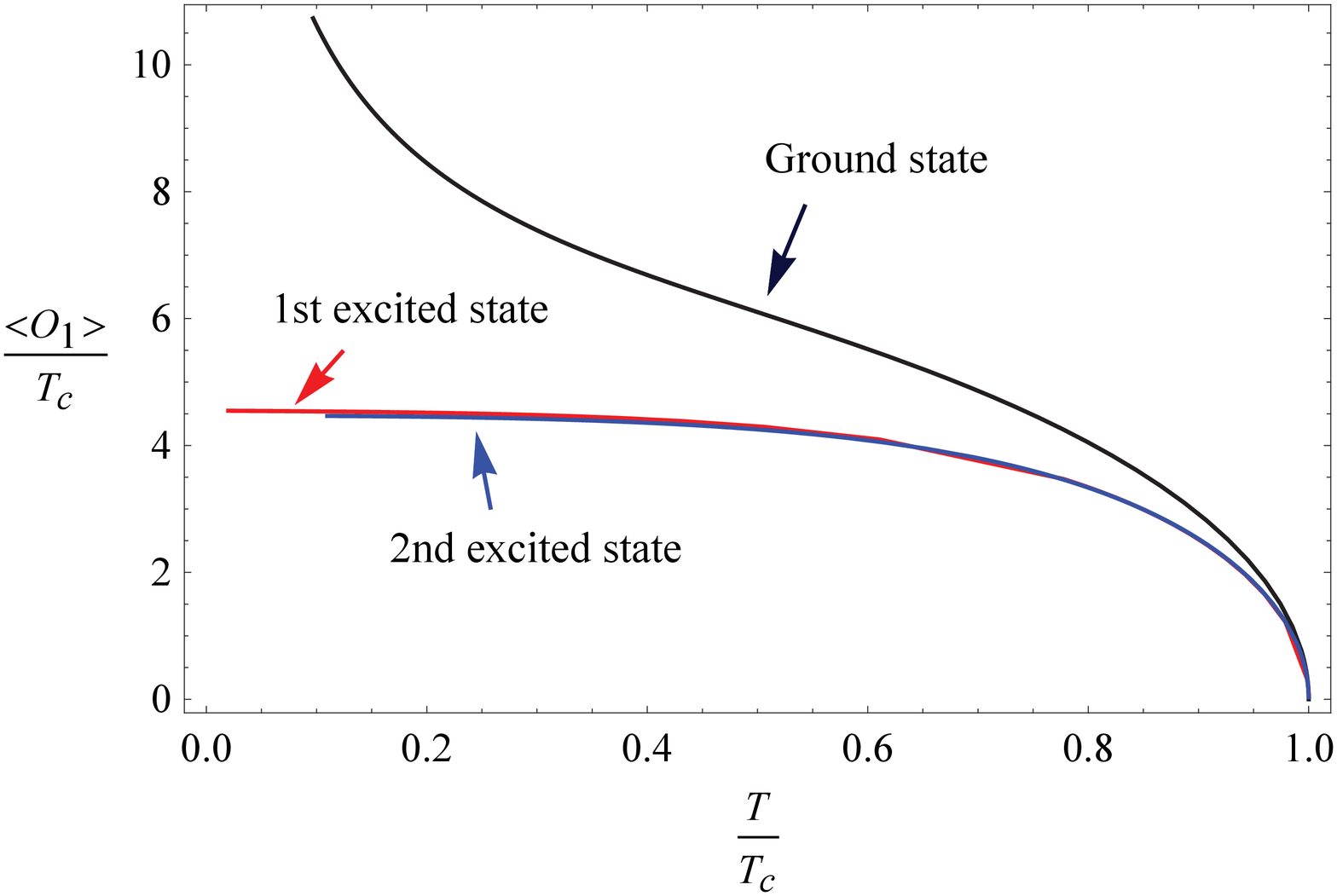,width=2.9in,angle=0,trim=0 0 0 0}
\epsfig{file=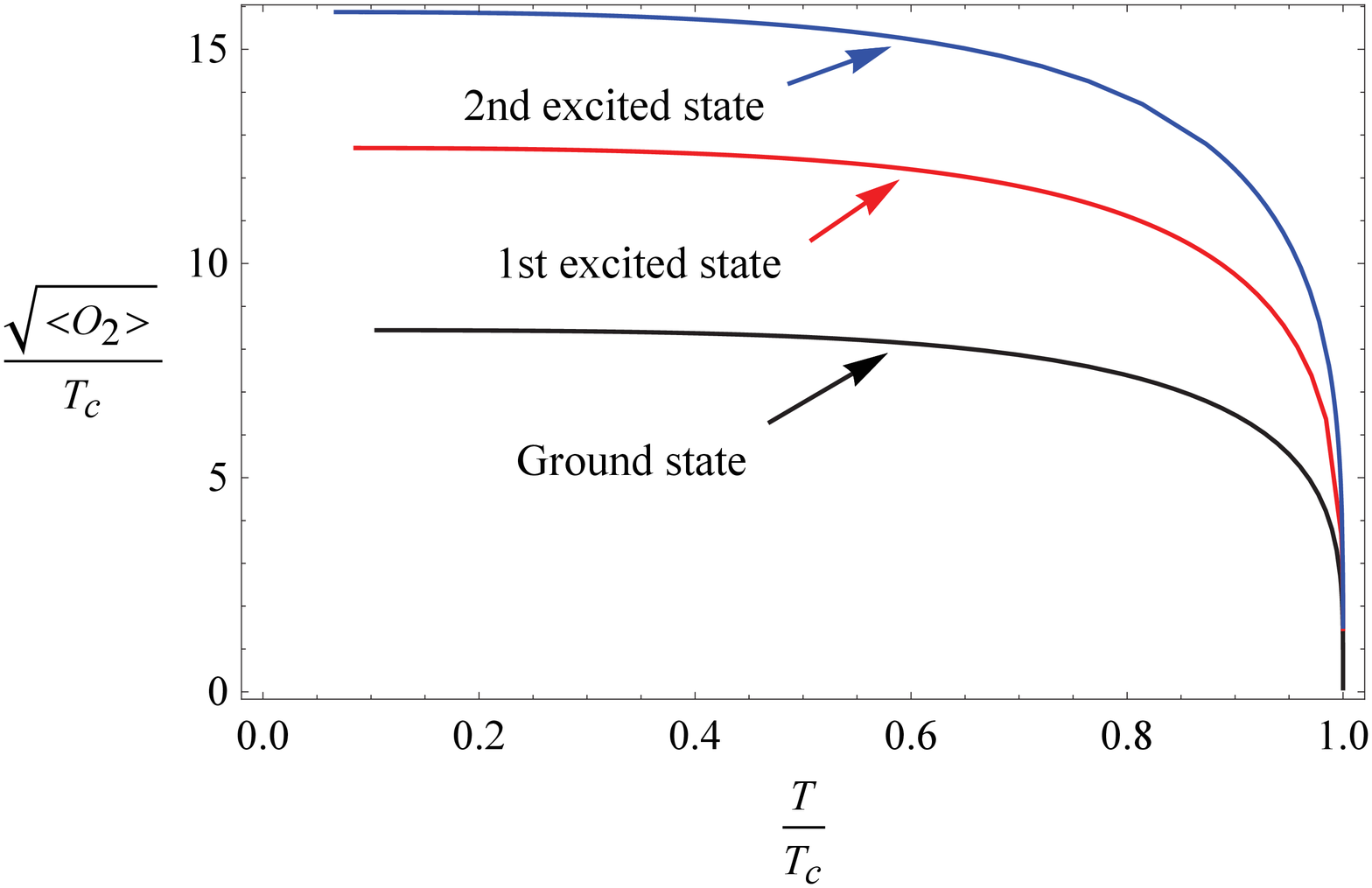,width=3.05in,angle=0,trim=0 0 0 0}
\end{center}
\caption{The condensate as a function of the
temperature for the two operators $\ocal_1$ (left) and $\ocal_2$ (right) of excited states. In both plots, the black,  red  and  blue lines denote ground state, first and second states, respectively.
  \label{fig:condensate}}
\end{figure}

 According to AdS/CFT duality,
 the expectation values of the  condensate operator $\langle\mathcal{O}_{i}\rangle$ is
dual to the scalar field $\psi^{(i)}$:
\be
\langle \ocal_i\rangle = \sqrt{2} \psi^{(i)} \,,\quad i=1,2.
\ee
 In Fig. \ref{fig:condensate}, we show the condensate as a function of the
temperature for the two operators $\ocal_1$ (left) and $\ocal_2$ (right) of excited states.
In both plots, the black,  red  and  blue lines denote ground state, first and second states, respectively.
 In the left panel,
the condensate of ground state for the operator $\langle\mathcal{O}_{1}\rangle$ appears
to diverge as the temperature $T\rightarrow0$, which means   the backreaction
of  the bulk metric can no longer be neglected. However, the condensate of excited states does not appear
to diverge as $T\rightarrow0$, and the condensate goes to a constant near zero temperature.  Moreover, the condensate of an excited state is smaller than the ground state.
Especially, we can see that  the value of the condensate for the second excited state  near zero temperature is very closed to that of the first excited state.
We have calculated  the condensate of zero temperature  until the sixth excited state, and find that this value  is about 4.4. In the right panel, the condensate of the ground
state for the operator $\mathcal{O}_{2}$  appears to converge as  the temperature $T\rightarrow0$, and the condensate of an excited state is larger than the ground state.

When the temperature drops below the critical temperature $T_c$, the model of holographic superconductor has a  second order phase transition
from  normal state to superconductor state. Moreover, there exists a square root behaviour near the  critical point, which could be expected from mean field theory.
We fit these condensation curves for the two operators $\ocal_1$ (left) and $\ocal_2$ (right) of excited states near $T\rightarrow T_c$ and present the fitting results as follows. For the  operator $\ocal_1$
\begin{displaymath}
\langle \ocal_1 \rangle\approx\left\{ \begin{array}{ll}
\, 9.3 \, T_c^{(0)} \, (1-T/T_c^{(0)})^{1/2} \,,  & \qquad\textrm{Ground state},\\
\, 8.9 \, T_c^{(1)} \, (1-T/T_c^{(1)})^{1/2} \,,  & \qquad\textrm{1st excited state}, \\
\, 8.5 \, T_c^{(2)} \, (1-T/T_c^{(2)})^{1/2} \,,  & \qquad\textrm{2nd excited state}.
 \end{array} \right.
\end{displaymath}
where the critical temperatures $T_c^{(0)} \approx 0.226 \rho^{1/2}$,  $T_c^{(1)} \approx 0.094 \rho^{1/2}$ and  $T_c^{(2)} \approx 0.07\rho^{1/2}$,
correspond to the ground, first  and second excited states, respectively. For the  operator $\ocal_2$
\begin{displaymath}
\langle \ocal_2 \rangle\approx\left\{ \begin{array}{ll}
144 \, (T_c^{(0)})^2 \, (1 - T/T_c^{(0)})^{1/2}\,,  & \qquad\textrm{Ground state},\\
329 \, (T_c^{(1)})^2 \, (1 - T/T_c^{(1)})^{1/2} \,,  & \qquad\textrm{1st excited state},\\
512 \, (T_c^{(2)})^2 \, (1 - T/T_c^{(2)})^{1/2} \,,  & \qquad\textrm{2nd excited state}.
 \end{array} \right.
\end{displaymath}
where  $T_c^{(0)} \approx 0.118 \rho^{1/2}$, $T_c^{(1)} \approx 0.079\rho^{1/2}$ and $T_c^{(2)} \approx 0.063\rho^{1/2}$ correspond to the ground, first  and second excited states, respectively. The above results come from fitting coefficients of condensation curves near critical point by exponential function with square root.
We can see that for the condensate $\langle \ocal_1 \rangle$, the fitting coefficients with higher excited state is smaller than ground state,
while, for the condensate $\langle \ocal_2 \rangle$, the fitting coefficients with higher excited state is larger than ground state.
It is noteworthy that the difference of the fitting coefficients between the consecutive states is  about $0.4$ for the condensate $\langle\mathcal{O}_{1}\rangle$ and $185$ for $\langle\mathcal{O}_{2}\rangle$.

From the above results,  it is clearly that because the excited state has a lower critical temperature than the ground
state,  the condensate of excited states is weaker  than the ground
state at the same temperature. That is, the Cooper pairs in excited states format more difficult than the ground state case.

\subsection{Conductivity}
In this subsection we study the conductivity in the excited states of holographic superconductors.
We turn on perturbations of  the vector potential $A_x$ in the bulk  geometry of Schwarzschild-AdS black hole.
 Considering the
Maxwell equation  with a time
dependence of $e^{- i \w t}$,  the linearized equations are given as
\be\label{eq:Axeq}
A_x'' + \frac{f'}{f} A_x' + \left(\frac{\w^2}{f^2} - \frac{2
\psi^2}{f} \right) A_x = 0 \,.
\ee
When imposing ingoing boundary conditions at the horizon,
we give the asymptotic behaviour of the Maxwell field at the boundary:
\be
A_x = A_x^{(0)} + \frac{A_x^{(1)}}{r} + \cdots.
\ee
According to the AdS/CFT duality and  Ohm's law we can obtain the conductivity
\be\label{eq:conductivity}
\sigma(\w) = - \frac{i A_x^{(1)}}{\w A_x^{(0)}} \,.
\ee
Because  the value  of the scalar $\psi$ in the excited stats is different from the ground state at the same temperature, the value of the conductivity  in Eq. (\ref{eq:Axeq}) could be changed with
the different excited states.

\begin{figure}[h]
\begin{center}
\epsfig{file=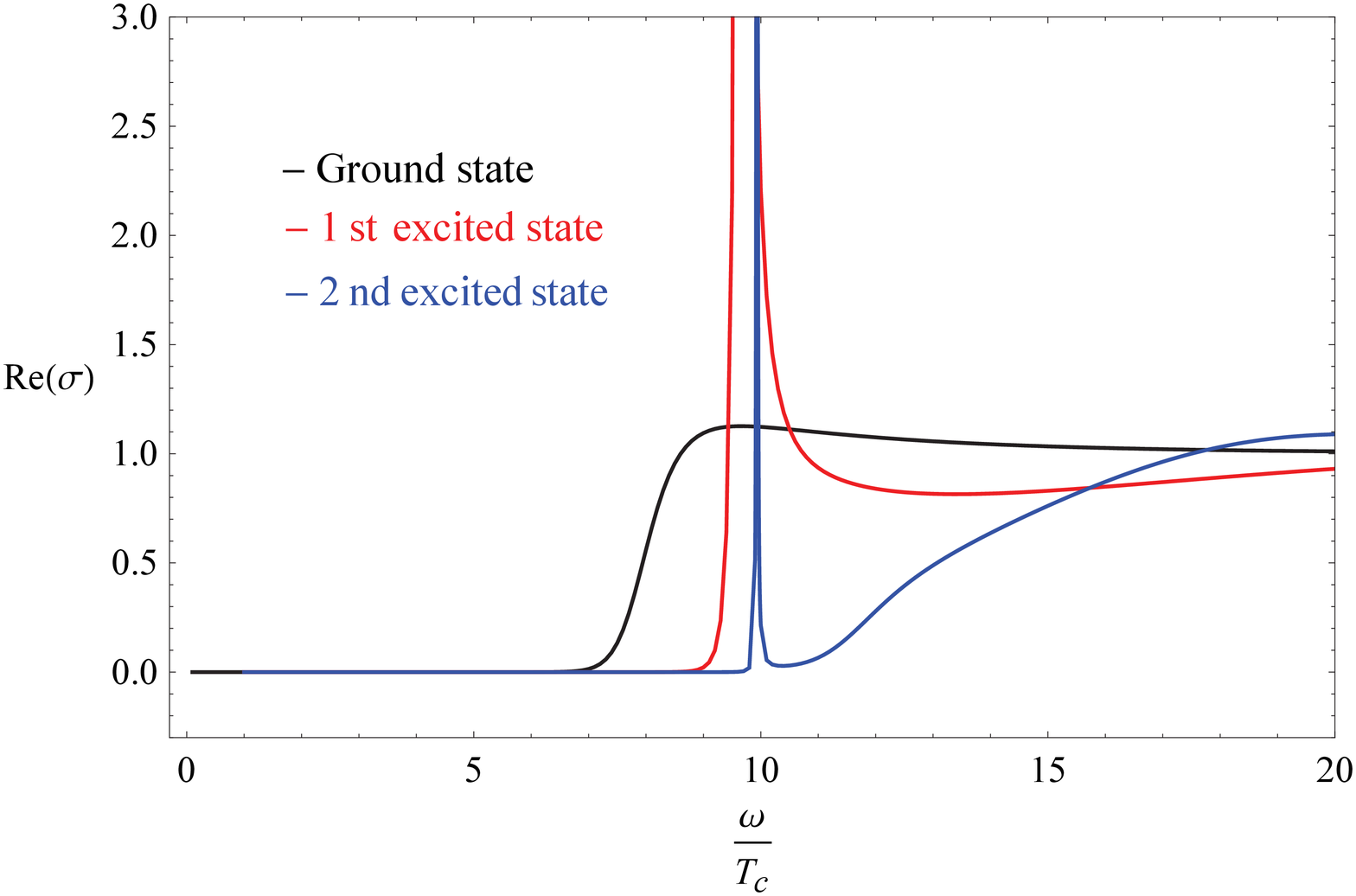,width=2.9in,angle=0,trim=0 0 0 0}%
\hspace{0.2cm}\epsfig{file=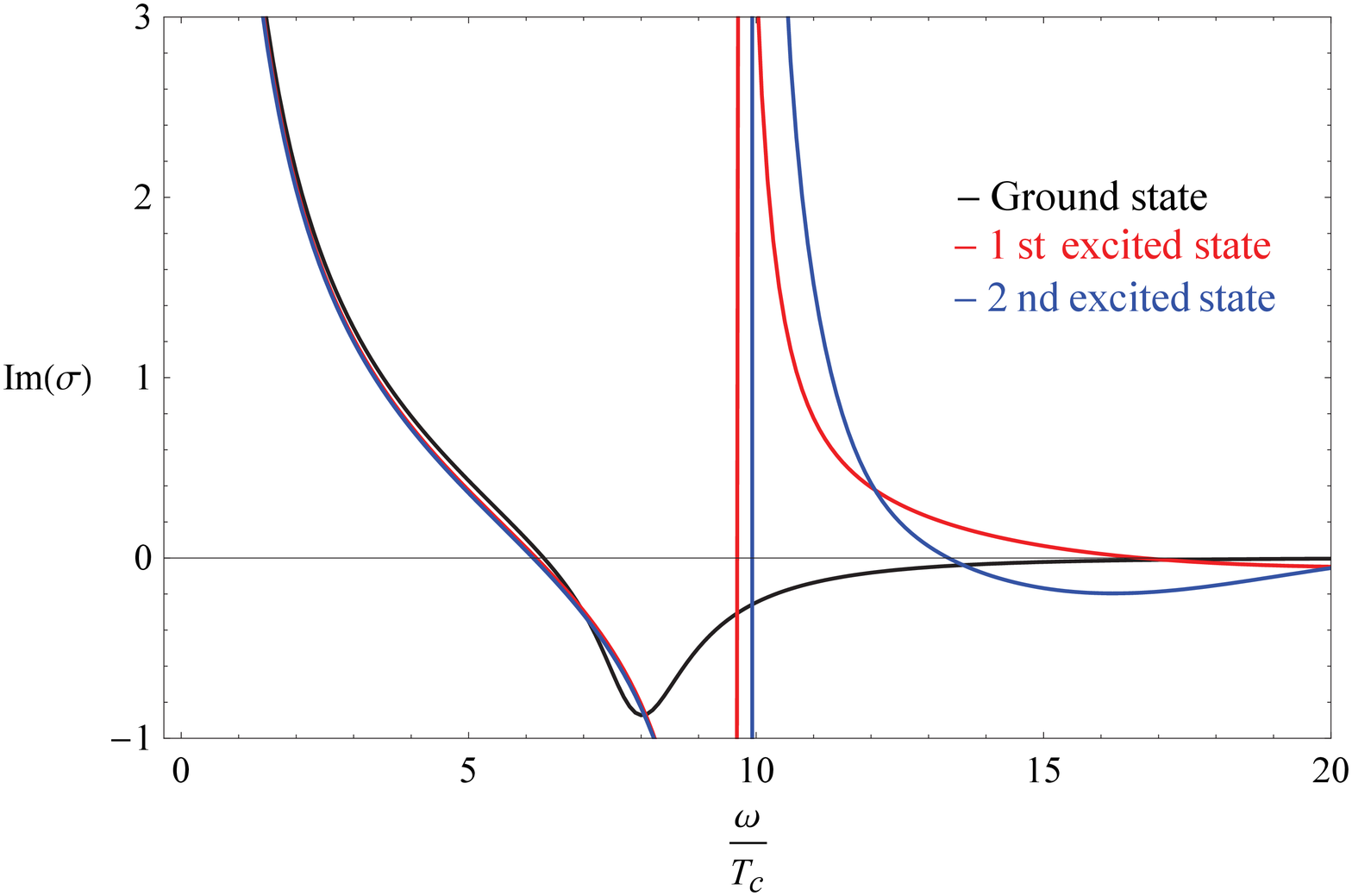,width=2.9in,angle=0,trim=0 0 0 0}%
\end{center}
\caption{ The real and imaginary parts of optical conductivity as a function of the frequency.
In both plots, the black,  red  and  blue lines denote ground state, first and second states, respectively.
  \label{fig:gap}}
\end{figure}

In Fig. \ref{fig:gap},  we plot the conductivity as a function of the
frequency  at the low temperature $T/T_c\simeq0.1$ for the  operators $\ocal_2$  of excited states.
 The left  and right panels correspond to  the real and imaginary parts of conductivity, respectively.
 In both plots, the black,  red  and  blue lines denote the ground state, first and second states, respectively.
 Similar to the ground state, there also  exists a  gap in the conductivity of the excited state.
  It is interesting to note that for the excited states, there exists an additional pole in
$\text{Im}[\sigma]$ and a delta function in $\text{Re}[\sigma]$  arising at low temperature inside the gap.
In addition, the form of  delta function of each higher excited states  in $\text{Re}[\sigma]$ has the narrower width.
In \cite{Horowitz:2008bn}, there exists the similar
configurations of the conductivity  as $m^2=-9/4$, and the authors pointed out that since holographic superconductors
are strongly coupled, there exist the   interactions
between  the quasiparticles excited
above the superconducting ground state, and the pole
 indicates that the interactions could be very strong and form
a bound state.
 Now, it is clear that these behaviors  are just the evidence of the  existence of the excited states of holographic superconductor.
Similar behaviours for conductivities were also found in  the type II Goldstone bosons \cite{Amado:2013xya}.
The authors related this behaviour of conductivities to the fact that these are not in the true ground state of the theory, and complemented the discussion by computing the quasinormal mode and confirm this picture. Furthermore, in \cite{Amado:2013lia} they computed the correct ground state.
So, these above behaviors of conductivities indicate the emergence of new bound states, which embodies the existence of excited states.

\section{Discussion}
In this paper, we reconsidered the static solutions of Einstein gravity coupled with Maxwell and a free, massive scalar field in four-dimensional AdS spacetime. Besides  the well-known ground state solution of holographic superconductor, we   constructed the numerical solutions of
holographic supercondecutor with excited states, including the case of
operators $\ocal_1$  and $\ocal_2$.
 Comparing with the ground
state solution in \cite{Hartnoll:2008vx}, we found that
when the temperature drops below the critical temperature $T_c^{(0)}$,  the condensate of holographic superconductor begin to appear, which could  be regarded as the ground state solution.
Further decreasing the temperature  to $T_c^{(1)}$, a new branch of solution with a node in radial direction begins to develop, which is the first excited state solution.
As the temperature continues to decrease to lower values, we  could obtain a series of excited states of holographic superconductor with differential critical temperatures.
It is well-known that the condensate of ground state for the operator $\langle\mathcal{O}_{1}\rangle$ appears
to diverge as the temperature $T\rightarrow0$. However, the condensate of excited states does not appear
to diverge and goes to a constant near zero temperature.  Moreover, the condensate of the excited state is smaller than the ground state.
Note that  the value of the condensate for the second excited state  near zero temperature is very closed to that of the first excited state.
We  calculated  the condensate $\langle\mathcal{O}_{1}\rangle$ near zero temperature  until the sixth excited state, and found that this value  is about 4.4.
For the  operator $\mathcal{O}_{2}$,  the condensate of an excited state is larger than the ground state.
We also studied the conductivity in the excited states of holographic superconductors.
 It is very interesting to mention that  each excited state has an additional pole in
$\text{Im}[\sigma]$ and a delta function in $\text{Re}[\sigma]$  arising at the low temperature inside the gap.
These
behaviors  are just the evidence of the  existence of the excited states.

Since in this paper we are using AdS/CFT to describe the superconductor with excite states, it is natural to ask what is the connection between holographic superconductor with excite states and  actual  superconductor matter in condensed physics. It is well known that if a thermodynamic system is in equilibrium state, it should be in the state with the minimum free energy. Holographic model of  superconductor investigated in \cite{Gubser:2009cg,Horowitz:2009ij} could correspond to the above state of minimum free energy. But for mesoscopic systems, such as nanostructures, due to the small size of the system and the small number of particles. The fluctuation of the system may make it run to a metastable state. Thus, these free energy surfaces are very complex and have many metastable states. The system can stay on these states for a long time.  In case of superconductors, it means that the sample should have one size at least of order the coherence length, and such superconductors are called mesoscopic superconductors \cite{Peeters2000}. In \cite{2002PhRvB66e4537V}, with the time-dependent Ginzburg-Landau equations, transitions between metastable states of a superconducting ring are investigated. So, excited states of holographic superconductor could be related to the metastable states of  mesoscopic superconductors.

There are many interesting extensions of our work. First,
we have noticed that  the value of the condensate $\langle\mathcal{O}_{1}\rangle$ near zero temperature  with higher excited state is about 4.4.
In addition, the difference of the critical chemical potential $\mu_{c}$ between the consecutive states is  about $5$ for both of  the condensates $\langle\mathcal{O}_{1}\rangle$ and $\langle\mathcal{O}_{2}\rangle$, but  the reason of these configurations is not
clear. It would be
very interesting to study these cases with  the semi-analytical method \cite{Siopsis:2010uq} to see how these values are related to excited states.
The second extension of our study is that away from the large charge limit, the backreaction of the matter fields
on the bulk metric needs to be included. Thus, one must solve the coupled differential equations
 including Einstein equations.
Finally, we are
planning to study the excited states of the p-wave and d-wave holographic superconductor
and construct the excited vector and tensor condensates in the background of black hole in
future work.

\subsection*{Acknowledgements}
We would like to thank Ying Zhong and Cheng-Long Jia   for  helpful discussion.  Parts of
 computations were performed on the   shared memory system at  institute of computational physics and complex systems in Lanzhou university. This work was supported by the National Natural Science Foundation of China (Grants
No. 11875151 and No. 11522541).

\end{document}